\documentclass{aa}  
\usepackage{graphicx}
\usepackage{txfonts}
\bibpunct{(}{)}{;}{a}{}{,} % to follow the A&A style

\begin{document} 

   \title{Planet-star interactions with precise transit timing}

   \subtitle{III. Entering the regime of dynamical tides\thanks{This research is partly based on (1) data obtained at the 1.5m telescope of the Sierra Nevada Observatory (Spain), which is operated by the Consejo Superior de Investigaciones Cient\'{\i}ficas (CSIC) through the Instituto de Astrof\'{\i}sica de Andaluc\'{\i}a, (2) observations collected with telescopes at the Rozhen National Astronomical Observatory, and (3) observations obtained with telescopes of the University Observatory Jena, which is operated by the Astrophysical Institute of the Friedrich-Schiller-University.}\thanks{The light curves will be available at the CDS.}}

   \author{G.~Maciejewski\inst{1}
          \and
          M.~Fern\'andez\inst{2}
          \and
          A.~Sota\inst{2}
          \and
          P.~J.~Amado\inst{2}
          \and
          D.~Dimitrov\inst{3}
          \and
          Y.~Nikolov\inst{3}
          \and
          J.~Ohlert\inst{4,5}
          \and
          M.~Mugrauer\inst{6}
          \and
          R.~Bischoff\inst{6} % richard.bischoff@uni-jena.de
          \and
          T.~Heyne\inst{6} % therese.heyne@uni-jena.de
          \and
          F.~Hildebrandt\inst{6} % felix.hildebrandt@uni-jena.de
          \and
          W.~Stenglein\inst{6} % wolfgang.stenglein@uni-jena.de
          \and
          A.~A.~Ar\'evalo\inst{7} % andytavo2@gmail.com
          \and
          S.~Neira\inst{7} % Ing.steve.neira@gmail.com
          \and
          L.~A.~Riesco\inst{7} % perlanivei@gmail.com
          \and
          V.~S\'anchez~Mart\'inez\inst{7} % vicki_mvsm@hotmail.com
          \and
          M.~M.~Verdugo\inst{7} % hfvirow@gmail.com
          }

   \institute{Institute of Astronomy, Faculty of Physics, Astronomy and Informatics,
              Nicolaus Copernicus University in Toru\'n, Grudziadzka 5, 87-100 Toru\'n, Poland,
              \email{gmac@umk.pl}
         \and
             Instituto de Astrof\'isica de Andaluc\'ia (IAA-CSIC), 
             Glorieta de la Astronom\'ia 3, 18008 Granada, Spain
         \and
             Institute of Astronomy and National Astronomical Observatory, 
             Bulgarian Academy of Sciences, 72 Tsarigradsko Chaussee Blvd.,
             1784, Sofia, Bulgaria
         \and
             Michael Adrian Observatorium, Astronomie Stiftung Trebur, 
             65428 Trebur, Germany
         \and
             University of Applied Sciences, Technische Hochschule Mittelhessen, 
             61169 Friedberg, Germany
         \and
             Astrophysikalisches Institut und Universit\"ats-Sternwarte, 
             Schillerg{\"a}sschen 2, 07745 Jena, Germany
         \and
             Valencia International University, Spain
             }
   \authorrunning{G.~Maciejewski et al.}
   \date{Received --- ; accepted --- }
 
% <casanova@iaa.es>
% <jmiguel@iaa.es>
 
  \abstract
   {Hot Jupiters on extremely short-period orbits are expected to be unstable to tidal dissipation and spiral toward their host stars. That is because they transfer the angular momentum of the orbital motion through tidal dissipation into the stellar interior. Although the magnitude of this phenomenon is related to the physical properties of a specific star-planet system, statistical studies show that tidal dissipation might shape the architecture of hot Jupiter systems during the stellar lifetime on the main sequence.}
   {The efficiency of tidal dissipation remains poorly constrained in star-planet systems. Stellar interior models show that the dissipation of dynamical tides in radiation zones could be the dominant mechanism driving planetary orbital decay. These theoretical predictions can be verified with the transit timing method.}
   {We acquired new precise transit mid-times for five planets. They were previously identified as the best candidates for which orbital decay might be detected. Analysis of the timing data allowed us to place tighter constraints on the orbital decay rate.}
   {No statistically significant changes in their orbital periods were detected for all five hot Jupiters in systems HAT-P-23, KELT-1, KELT-16, WASP-18, and WASP-103. For planets HAT-P-23~b, WASP-18~b, and WASP-103~b, observations show that the mechanism of the dynamical tides dissipation probably does not operate in their host stars, preventing them from rapid orbital decay. This finding aligns with the models of stellar interiors of F-type stars, in which dynamical tides are not fully damped due to convective cores. For KELT-16~b, the span of transit timing data was not long enough to verify the theoretical predictions. KELT-1~b was identified as a potential laboratory for studying the dissipative tidal interactions of inertial waves in a convective layer. Continued observations of those two planets may provide further empirical verification of the tidal dissipation theory.}
   {}


   \maketitle
%
%-------------------------------------------------------------------

\section{Introduction}\label{Sect:Intro}

The statistical studies of the planetary systems harbouring hot Jupiters provide substantial evidence of dissipative tidal interactions between those massive planets and their host stars. In a typical configuration, in which the host star rotates slower than its close planetary companion orbits it, the angular momentum of the orbital motion is transferred into the stellar spin due to tidal dissipation in the stellar interior. The population of hot Jupiter host stars was found to have a lower Galactic velocity dispersion than the field stars in a reference sample \citep{2019AJ....158..190H}. This kinematical youth suggests that their planets must spiral in due to tidal interactions in time scales noticeably shorter than the stellar evolution on the main sequence. Furthermore, the host stars with close-orbiting giant planets tend to be younger in gyro-chronological dating compared to their ages determined from stellar-evolutionary models \citep{2014MNRAS.442.1844B,2015A&A...577A..90M,2021ApJ...919..138T}. They are supposed to rotate faster because their spiralling-in planets have spun them up.

The magnitude of tidal dissipation in a star is quantified by a modified tidal quality factor defined as 
\begin{equation}
  Q'_{\star} = \frac{2 \pi E_{\rm{tide}}}{\oint D \rm{d} t} \frac{3}{2 k_2} \, , \;
\end{equation}
where $E_{\rm{tide}}$ is the maximum energy stored in the tide, $D$ is the dissipation integrated over the tidal period $P_{\rm{tide}}$, and $k_2$ is the second-order potential Love number \citep{2020MNRAS.498.2270B}. The tidal period is related to the stellar rotation period $P_{\star}$ and planetary orbital period $P_{\rm {orb}}$ with the formula
\begin{equation}
  \frac{1}{P_{\rm{tide}}} = 2 \left( \frac{1}{P_{\rm {orb}}} - \frac{1}{P_{\star}} \right) \, . \;
\end{equation}
The value of $Q'_{\star}$ encompasses the physical properties of the specific star-planet system. Hence it might significantly vary from one system to another and makes us take the results of population-wide studies of hot-Jupiters as a rather rough approximation \citep{2020MNRAS.498.2270B}.

The stellar interior might dissipate the tidal energy under the equilibrium (EQ) and dynamical regimes. In the former, the global-scale flow is induced by the hydrostatic response of the stellar figure over the planet's gravitational potential. Those tides are dissipated in a convective layer\footnote{The contribution of dissipation in a convective core is predicted to be negligibly small \citep{2020MNRAS.498.2270B}}. The efficiency of this mechanism is, however, low for main-sequence stars, resulting in its component of the modified tidal quality factor of $Q'_{\star,\rm{EQ}}>10^{10}$ \citep{2020MNRAS.498.2270B}. In the dynamical regime, internal gravity waves (IGW) in radiation layers or inertial waves (IW) in convective layers are excited in response to tidal forcing. Calculations show that these mechanisms can be efficient in tidal dissipating under favourable conditions in which non-adiabatic or non-linear effects can operate. Dissipation of internal gravity waves in radiation layers might be substantially enhanced, resulting in $Q'_{\star,\rm{IGW}}$ of the order of $10^6$ for main-sequence F-type stars, and even as low as $10^{5}-10^6$ for 0.4 $M_{\odot}$ at the same evolutionary stage \citep{2020MNRAS.498.2270B}. Dissipation due to inertial waves in convective layers only operates if $P_{\rm{tide}} > 2 P_{\star}$, and might yield $Q'_{\star,\rm{IW}}$ as low as $10^{5}-10^6$ for fast-rotating stars with masses below 1.1 $M_{\odot}$ \citep{2020MNRAS.498.2270B}. Dissipation of the equilibrium tides seems to be too weak to have observable effects in individual planetary systems. On the other hand, the magnitude of dynamical tides dissipation could manifest as orbital decay detectable in decadal timescales for favourable systemic configurations. 

In our previous papers \citep{2018AcA....68..371M,2020AcA....70....1M}, we used the transit timing method to probe values of $Q'_{\star}$ for a sample of systems with massive planets on extremely tight orbits. We selected them among the candidates for which orbital decay due to tidal dissipation might be detected over a decade if their modified tidal quality factors were $10^6$. In this study, we extend the time coverage of observations for five systems HAT-P-23 \citep{2011AJ....142...84F}, KELT-1 \citep{2012ApJ...761..123S}, KELT-16 \citep{2017AJ....153...97O}, WASP-18 \citep{2009Natur.460.1098H}, and WASP-103 \citep{2014A&A...562L...3G} using both high-quality ground-based follow-up observations and photometric time series from the Transiting Exoplanet Survey Satellite \citep[TESS,][]{2014SPIE.9143E..20R}. The data sets with homogeneously determined mid-transit times allowed us to place tighter constraints on the values of $Q'_{\star}$ in these systems and explore the dynamical tides' regime.

\section{Observations and data reduction}\label{Sect:Obs}

\subsection{Ground-based observations}\label{SSect:GObs}

We acquired ten transit light curves for HAT-P-23~b, five for KELT-1~b, ten for KELT-16~b, and eight for WASP-103~b (including two light curves of the transit observed on 2019 June 01). We employed six instruments: the 2.0 m Ritchey-Chr\'etien-Coud\'e telescope (Rozhen) at the National Astronomical Observatory Rozhen (Bulgaria) equipped with a Roper Scientific \mbox{Vers}Array 1300B CCD camera; the 1.5 m Ritchey-Chr\'etien telescope (OSN150) at the Sierra Nevada Observatory (OSN, Spain) with a Roper Scientific VersArray 2048B CCD camera; the 1.2 m Trebur one-meter telescope (Trebur) at the Michael Adrian Observatory in Trebur (Germany) with an SBIG STL-6303 CCD camera; the 0.9 m Ritchey-Chr\'etien telescope (OSN90) at OSN with a Roper Scientific VersArray 2048B CCD camera; the 0.9/0.6 m Schmidt Teleskop Kamera \citep[Jena,][]{2010AN....331..449M} at the University Observatory Jena (Germany); and the 0.6 m Cassegrain photometric telescope (Torun) at the Institute of Astronomy of the Nicolaus Copernicus University in Toru\'n (Poland) with an FLI 16803 CCD camera. 

The telescope defocusing technique, in which the stellar point spread function is broadened, spreading starlight over many CCD pixels, was used at each instrument. It reduces flat-fielding errors and minimises the amount of observing time lost for CCD readout \citep[e.g.][]{2009MNRAS.396.1023S}. The observations were primarily performed without any filter to maximise the signal-to-noise ratio for precise transit timing, and only occasionally were the observations acquired through an $R$-band filter. The only exception is KELT-1, the brightest star in our sample ($G \approx 10.6$ mag). For that field, photometric time series were secured with the $R$ filter to avoid saturation of the target and comparison stars. If available, auto-guiding was applied to minimise field drifts during each run. Otherwise, tracking corrections were applied manually to keep stellar images around the fixed position in a CCD matrix within $\approx 5 \arcsec$.

The observations were scheduled to secure 60-90 minutes of out-of-transit monitoring before and after each transit to remove trends reliably. For several light curves, data portions were lost due to unfavourable weather conditions or observing constraints. Details on the individual observing runs are collected in Table~\ref{tab.Obs}.

%------------------ TABLE OBS
\begin{table*}[h]
\caption{Details on the observing runs.} 
\label{tab.Obs}      
\centering                  
\begin{tabular}{l l c c c c c c c}      
\hline\hline                
Date UT (Epoch)  & Telescope & Band & UT start--end  &  $X$                                & $N_{\rm{obs}}$ & $t_{\rm{exp}}$ (s) & $\Gamma$ & pnr (ppth)\\
\hline
\multicolumn{9}{c}{HAT-P-23 b} \\
2019 Sep 17 (3209) & Jena   & clear & 20:08--23:31 & $1.21 \rightarrow 1.73$                  & 211 & 45 & 1.05 & 1.51 \\
2019 Oct 15 (3232) & Jena   & clear & 17:43--20:52 & $1.21 \rightarrow 1.49$                  & 144 & 45 & 1.07 & 2.13 \\
2020 Jul 21 (3463) & Trebur & clear & 21:17--01:56 & $1.41 \rightarrow 1.19 \rightarrow 1.34$ & 240 & 60 & 0.87 & 0.93 \\
2020 Aug 07 (3477) & OSN150 & R     & 20:32--01:15 & $1.40 \rightarrow 1.07 \rightarrow 1.16$ & 761 & 20 & 2.68 & 0.99 \\
2020 Oct 14 (3533) & OSN150 & clear & 18:57--00:00 & $1.07 \rightarrow 2.60$                  & 710 & 20 & 2.35 & 1.58 \\
2021 Jul 04 (3750) & OSN90  & clear & 23:15--03:27 & $1.28 \rightarrow 1.07 \rightarrow 1.16$ & 422 & 30 & 1.67 & 0.91 \\
2021 Jul 09 (3754) & Rozhen & clear & 20:41--00:15 & $1.40 \rightarrow 1.10 \rightarrow 1.12$ & 458 & 20 & 2.39 & 0.70 \\
2021 Jul 15 (3759) & OSN90  & clear & 21:41--02:24 & $1.52 \rightarrow 1.07 \rightarrow 1.12$ & 474 & 30 & 1.67 & 1.03 \\
2021 Aug 01 (3773) & OSN90  & clear & 22:07--01:46 & $1.17 \rightarrow 1.07 \rightarrow 1.17$ & 285 & 40 & 1.31 & 1.22 \\
2021 Sep 04 (3801) & OSN150 & clear & 20:09--00:56 & $1.14 \rightarrow 1.07 \rightarrow 1.47$ & 677 & 20 & 2.35 & 0.65 \\
\multicolumn{9}{c}{KELT-1 b} \\
2020 Oct 11 (2649) & OSN90  & R     & 20:21--00:56 & $1.15 \rightarrow 1.00 \rightarrow 1.10$ &  457 & 30 & 1.67 & 0.96\\
2020 Oct 28 (2663) & OSN90  & R     & 20:36--02:08 & $1.03 \rightarrow 1.00 \rightarrow 1.57$ &  549 & 30 & 1.67 & 0.82\\
2021 Sep 01 (2916) & OSN150 & R     & 21:59--02:49 & $1.32 \rightarrow 1.00 \rightarrow 1.04$ & 1002 & 15 & 3.47 & 0.80\\
2021 Sep 29 (2939) & OSN90  & R     & 21:42--02:31 & $1.09 \rightarrow 1.00 \rightarrow 1.20$ &  481 & 30 & 1.67 & 0.86\\
2021 Oct 10 (2948) & OSN90  & R     & 20:42--01:34 & $1.12 \rightarrow 1.00 \rightarrow 1.16$ &  242 & 30 & 1.67 & 0.79\\
\multicolumn{9}{c}{KELT-16 b} \\
2019 Aug 14 (1510) & Toru\'n & clear & 20:05--00:28 & $1.17 \rightarrow 1.07 \rightarrow 1.18$ & 508 & 27 & 2.00 & 1.48 \\
2020 Jul 22 (1864) & OSN90   & clear & 21:00--01:22 & $1.57 \rightarrow 1.00 \rightarrow 1.01$ & 385 & 35 & 1.47 & 1.08 \\
2020 Aug 20 (1894) & Trebur  & clear & 22:36--03:01 & $1.05 \rightarrow 1.67$                  & 257 & 50 & 1.02 & 0.87 \\
2020 Aug 21 (1895) & OSN90   & clear & 21:24--01:45 & $1.08 \rightarrow 1.00 \rightarrow 1.19$ & 337 & 40 & 1.31 & 0.70 \\
2020 Aug 22 (1896) & OSN90   & clear & 20:44--01:01 & $1.15 \rightarrow 1.00 \rightarrow 1.10$ & 336 & 40 & 1.31 & 0.81 \\
2020 Aug 23 (1897) & OSN90   & clear & 19:48--00:19 & $1.30 \rightarrow 1.00 \rightarrow 1.05$ & 353 & 40 & 1.31 & 0.72 \\
2021 Aug 30 (2281) & OSN90   & clear & 22:05--03:07 & $1.01 \rightarrow 1.00 \rightarrow 1.75$ & 678 & 20 & 2.31 & 1.02 \\
2021 Aug 31 (2282) & Trebur  & clear & 22:09--02:08 & $1.06 \rightarrow 1.60$                  & 244 & 50 & 1.02 & 0.78 \\
2021 Sep 02 (2284) & OSN150  & R     & 20:00--00:51 & $1.15 \rightarrow 1.00 \rightarrow 1.17$ & 639 & 25 & 2.20 & 0.90 \\
2021 Oct 04 (2317) & OSN90   & clear & 19:39--00:52 & $1.01 \rightarrow 1.00 \rightarrow 1.78$ & 320 & 40 & 1.31 & 1.01 \\
\multicolumn{9}{c}{WASP-103 b} \\
2019 May31 (2351) & OSN150 & clear & 22:17--03:36 & $1.31 \rightarrow 1.15 \rightarrow 1.72$ & 422 & 40 & 1.32 & 0.84\\
2019 Jun 01 (2352) & OSN150 & clear & 20:55--02:30 & $1.69 \rightarrow 1.15 \rightarrow 1.38$ & 398 & 40 & 1.19 & 0.92\\
                  & Jena   & clear & 21:31--00:27 & $1.51 \rightarrow 1.31 \rightarrow 1.45$ & 183 & 45 & 1.05 & 1.75\\
2019 Jun 26 (2379) & Trebur & clear & 21:14--01:07 & $1.37 \rightarrow 1.36 \rightarrow 1.99$ & 195 & 55 & 0.95 & 1.39\\
2019 Jun 27 (2380) & Rozhen & clear & 19:17--23:19 & $1.28 \rightarrow 1.21 \rightarrow 1.55$ & 463 & 26 & 1.93 & 0.86\\
2021 May 21 (3130) & OSN150 & R     & 23:45--03:15 & $1.20 \rightarrow 1.15 \rightarrow 1.39$ & 230 & 50 & 1.15 & 1.82\\
2021 Jun 03 (3144) & OSN150 & R     & 21:44--00:49 & $1.37 \rightarrow 1.15 \rightarrow 1.18$ & 205 & 50 & 1.15 & 1.24\\
2021 Jul 12 (3186) & Rozhen & clear & 19:01--23:16 & $1.23 \rightarrow 1.21 \rightarrow 1.97$ & 597 & 20 & 2.39 & 0.71\\
\hline                                   
\end{tabular}
\tablefoot{Date UT is given for the beginning of an observing run. Epoch is the transit number from the initial ephemeris given in the discovery papers. $X$ tracks the target's airmass during a run. $N_{\rm{obs}}$ is the number of useful scientific exposures. $t_{\rm{exp}}$ is the exposure time used. $\Gamma$ is the median number of exposures per minute. $pnr$ is the photometric noise rate \citep{2011AJ....142...84F} in parts per thousand (ppth) of the normalised flux per minute of observation.}
\end{table*}
%------------------ END OF TABLE

\begin{figure}
	\includegraphics[width=\columnwidth]{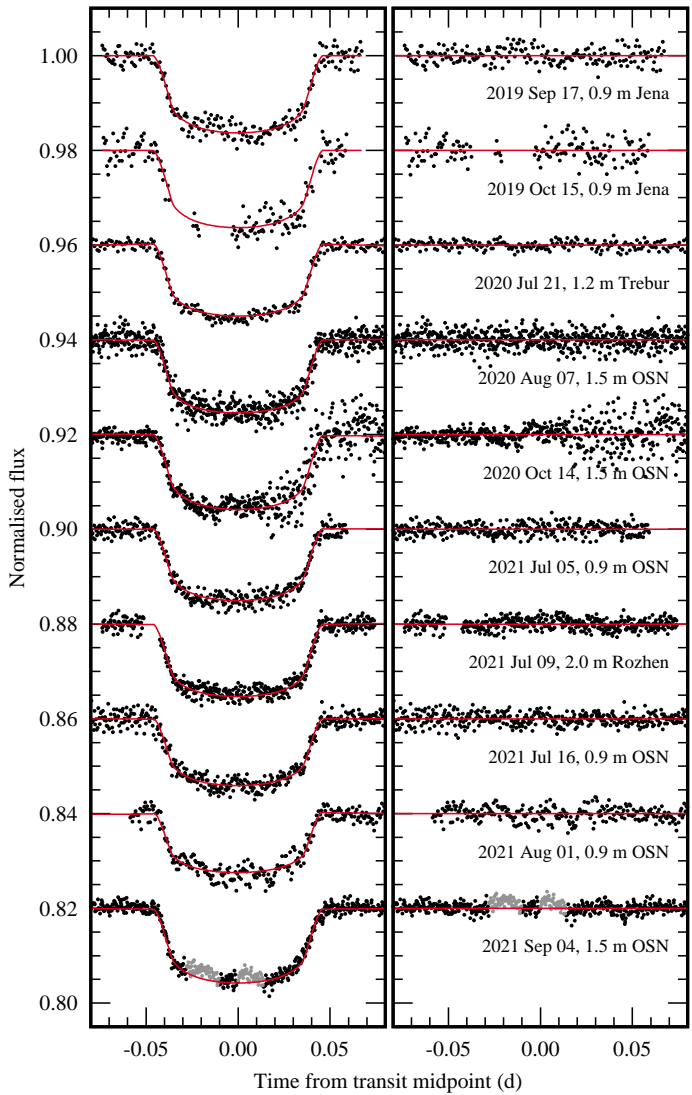}
    \caption{Left: new transit light curves for HAT-P-23~b, sorted by the observation date. The best-fitting model is drawn with red lines. A signature of star-spot occultation identified in the light curve acquired on 2021 September 04 is marked with grey points. These measurements were masked out in the transit modelling. Right: photometric residuals from the transit model.}
    \label{fig:h23lcs}
\end{figure}

\begin{figure}
	\includegraphics[width=\columnwidth]{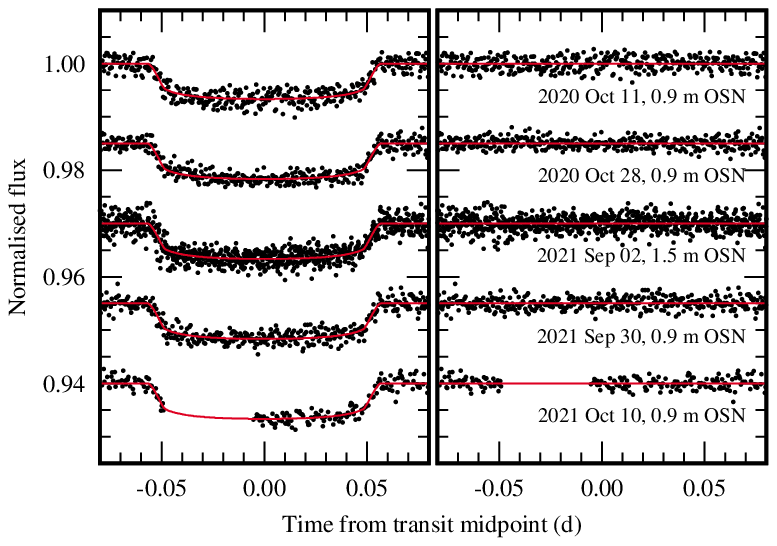}
    \caption{The same as Fig.~\ref{fig:h23lcs} but for KELT-1~b.}
    \label{fig:k01lcs}
\end{figure}

\begin{figure}
	\includegraphics[width=\columnwidth]{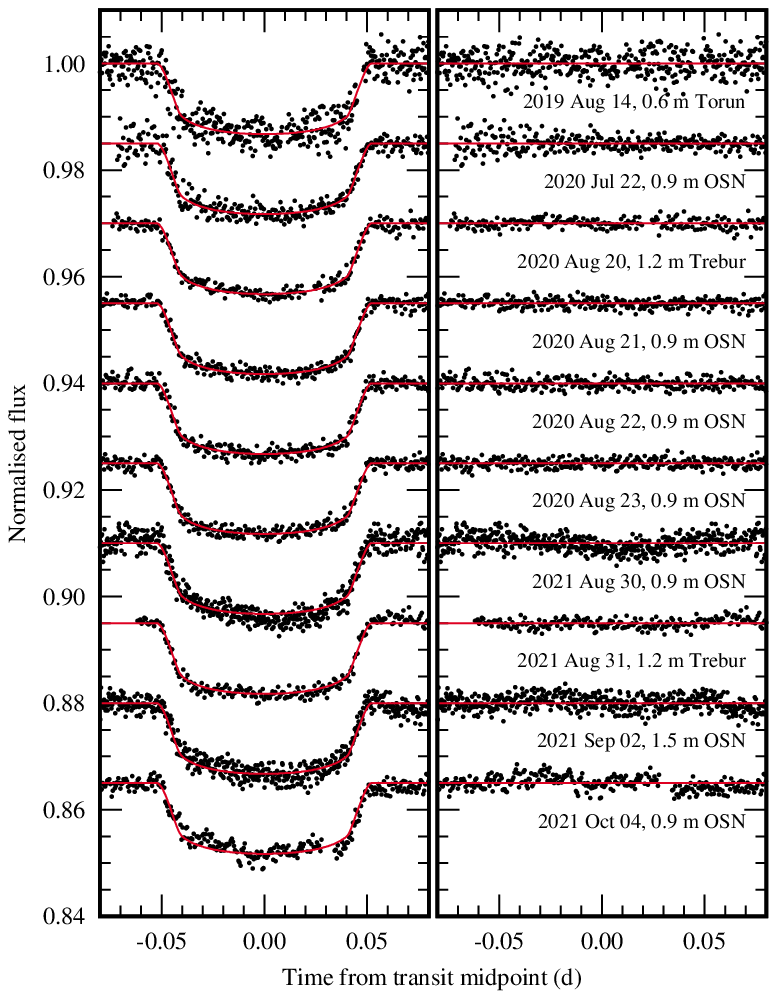}
    \caption{The same as Fig.~\ref{fig:h23lcs} but for KELT-16~b.}
    \label{fig:k16lcs}
\end{figure}

\begin{figure}
	\includegraphics[width=\columnwidth]{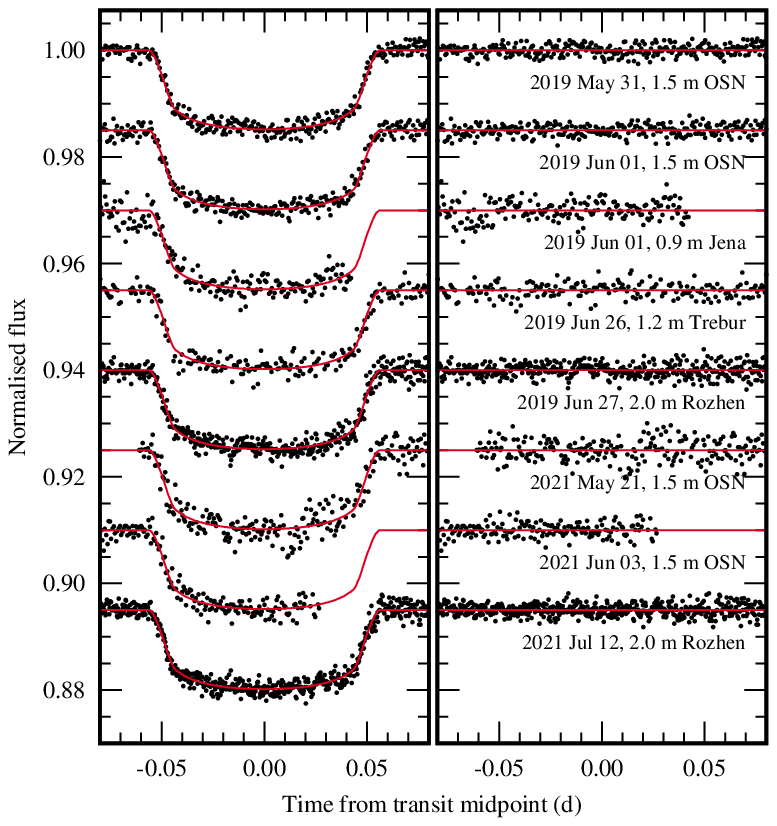}
    \caption{The same as Fig.~\ref{fig:h23lcs} but for WASP-103~b.}
    \label{fig:w103lcs}
\end{figure}

AstroImageJ \citep{2017AJ....153...77C} was used for data processing and photometric extraction of the final light curves. The science frames were preprocessed following a standard procedure, including de-biasing or dark-current correction and flat-fielding with sky flat-field frames. The timestamps of mid-exposures were transformed into barycentric Julian dates and barycentric dynamical time $\rm{BJD_{TDB}}$ using a built-in converter. Fluxes were obtained with the aperture photometry method with the aperture size and ensemble of comparison stars optimised in trial iterations. Then, normalisation to unity outside the transit was performed simultaneously with a trial transit model and de-trending against airmass, time, and seeing\footnote{Thanks to precise guiding, trends against the X-Y position on a CCD matrix were neglectable.}. The final light curves are plotted in Figs.~\ref{fig:h23lcs}--\ref{fig:w103lcs}.

\subsection{TESS observations}\label{SSect.TESSobs}

Three systems of our sample, KELT-1, KELT-16, and WASP-18, were observed with the Transiting Exoplanet Survey Satellite \citep[TESS,][]{2014SPIE.9143E..20R} in a 2-minute cadence mode. The photometric time series were extracted from Pre-search Data Conditioning Simple Aperture Photometry which is available via the exo.MAST portal\footnote{https://exo.mast.stsci.edu}. The details on individual observing runs are given in Table~\ref{tab.ObsTESS}.

%------------------ TABLE OBS.TESS
\begin{table*}[h]
\caption{Details on the TESS observations used.} 
\label{tab.ObsTESS}      
\centering                  
\begin{tabular}{l c c c c c c}      
\hline\hline                
System & Sector  & Camera & from -- to (UT) & $N_{\rm{obs}}$ & pnr (ppth) & $N_{\rm{tr}}$ \\
\hline
KELT-1  & 17 & 2 & 2019 Oct 07 -- 2019 Nov 02 & 12892 & $1.67$ & 14 \\
KELT-16 & 15 & 1 & 2019 Aug 15 -- 2019 Sep 11 & 13129 & $3.76$ & 18 \\
        & 41 & 1 & 2021 Jul 23 -- 2021 Aug 20 & 18322 & $4.12$ & 26 \\
WASP-18 & 29 & 2 & 2020 Aug 26 -- 2020 Sep 22 & 14361 & $0.79$ & 21 \\
        & 30 & 2 & 2020 Sep 22 -- 2020 Oct 21 & 16573 & $0.78$ & 24 \\
\hline                                   
\end{tabular}
\tablefoot{$N_{\rm{obs}}$ is the number of useful data points. $pnr$ is the photometric noise rate. $N_{\rm{tr}}$ is the number of complete transit light curves used in this study.}
\end{table*}
%------------------ END OF TABLE

\begin{figure}
	\includegraphics[width=\columnwidth]{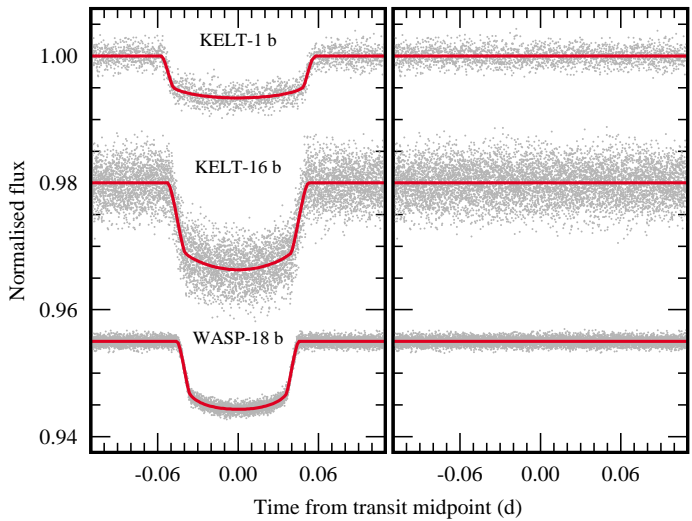}
    \caption{Left: phase folded TESS transit light curves for KELT-1~b, KELT-16~b, and WASP-18~b. The best-fitting models are plotted with red lines. Right: photometric residuals from the transit models.}
    \label{fig:tesslcs}
\end{figure}

The Savitzky-Golay filter implemented in the Lightkurve package \citep[ver. 2.0,][]{2018ascl.soft12013L} was used to remove any low-frequency trends from astrophysical or systematic effects on time scales much longer than the expected transit duration. Data points falling in transits and occultations were masked out using preliminary transit ephemerides. Since the durations of the transits were 2.1-2.7 hours, the length of the filter window was set to 6 hours. Measurements with fluxes or flux errors listed as NaN were automatically removed. Apparent outliers were identified by visual inspection and then rejected. The light curves of complete transits within some time margins before and after each transit were extracted for further analysis. The length of these margins was set as twice the transit duration. The phase-folded transit light curves are shown in Fig.~\ref{fig:tesslcs}.

\section{Results}\label{Sect:Results}

\subsection{Transit light curve modeling}\label{SSect:Results.Parameters}

The Transit Analysis Package \citep[TAP,][]{2012AdAst2012E..30G} was employed to model the transit light curves. For each planet, the orbital inclination $i_{\rm{b}}$, the semi-major axis scaled in stellar radii $a_{\rm{b}}/R_{\star}$, the ratio of planet to star radii $R_{\rm{b}}/R_{\star}$, the limb darkening coefficients (LDCs) of a quadratic law $u_1$ and $u_2$, times of transit midpoints $T_{\rm{mid}}$, and possible flux variations approximated with a second-order polynomial were fitted. The systemic parameters were linked for all light curves. In test runs, we searched for variation in $R_{\rm{b}}/R_{\star}$ in the individual bands, but we found no statistically significant differences. The LDCs for light curves acquired in the same band were also linked. We obtained ground-based and TESS light curve solutions separately in the trial iterations and compared the results. We found no statistically significant differences between these datasets and between the results reported in \citet{2018AcA....68..371M} and \citet{2020AcA....70....1M}. Thus, in the final iterations, the best-fitting models were found using joint data sets from \citet{2018AcA....68..371M} and \citet{2020AcA....70....1M} and the groud-based and TESS light curves which are presented in this paper. This approach allowed us to refine most of the parameters with higher precision. Our tests showed that the greater uncertainties for $R_{\rm{b}}/R_{\star}$ in the HAT-P-23, KELT-1, and WASP-103 systems are caused by allowing $u_1$ and $u_2$ to be the free parameters. This approach is advocated by \citet{2022AJ....163..228P} who have found that discrepancies between the predicted and determined values of TESS LDCs can reach up to $\Delta u \approx 0.2$. The best-fitting parameters and their uncertainties were taken as 50, 15.9, and 84.1 percentiles of the marginalised posterior probability distributions generated by ten Markov chains Monte Carlo (MCMC), each $10^6$ steps long with a 10\% burn-in phase. The results are collected in Table~\ref{tab.res} together with some recent literature values for easy comparison. The new values of $T_{\rm{mid}}$ are given in Table~\ref{tab.TTimes} in Appendix~\ref{App:1}. The models are sketched in Figs.~\ref{fig:h23lcs}--\ref{fig:w103lcs} and Fig.~\ref{fig:tesslcs} together with the residuals.

\begin{table*}[h]
\caption{Systemic parameters refined from transit light curves.} 
\label{tab.res}      
\centering                  
\begin{tabular}{c c c c c c l}      
\hline\hline                
$i_{\rm{b}}$ $(^{\circ})$ & $a_{\rm{b}}/R_{\star}$ & $R_{\rm{b}}/R_{\star}$ & $u_1$, $u_2$ $(R)$ & $u_1$, $u_2$ (clear) & $u_1$, $u_2$ (TESS) & Source \\
\hline
\multicolumn{7}{c}{HAT-P-23 b} \\
$84.45^{+0.44}_{-0.39}$ & $4.405^{+0.050}_{-0.049}$ & $0.11647^{+0.00076}_{-0.00079}$ & $0.32^{+0.10}_{-0.10}$, $0.16^{+0.17}_{-0.17}$ & $0.54^{+0.09}_{-0.09}$, $-0.08^{+0.16}_{-0.15}$ & -- & this paper \\
$85.23^{+0.54}_{-0.48}$ & $4.465^{+0.062}_{-0.059}$ & $0.11612^{+0.00058}_{-0.00060}$ & $0.35^{\rm {a}}$, $0.29^{\rm {a}}$ & $0.40^{\rm {a}}$, $0.28^{\rm {a}}$ & -- & Mac18 \\
\multicolumn{7}{c}{KELT-1 b} \\
$85.2^{+2.5}_{-1.8}$    & $3.58^{+0.11}_{-0.12}$    & $0.0766^{+0.0008}_{-0.0008}$ & $0.27^{+0.15}_{-0.14}$, $0.24^{+0.22}_{-0.24}$ & $0.46^{+0.15}_{-0.16}$, $-0.03^{+0.24}_{-0.23}$ & $0.42^{+0.15}_{-0.15}$, $-0.06^{+0.23}_{-0.23}$ & this paper \\
$85.3^{+2.9}_{-2.6}$ & $3.55^{+0.12}_{-0.18}$ & $0.0762^{+0.0012}_{-0.0010}$ & $0.27^{\rm {a}}$, $0.33^{\rm {a}}$ & -- & -- & Mac18 \\
$85.8^{+2.7}_{-2.8}$ & $3.59^{+0.10}_{-0.18}$ & $0.07612^{+0.00095}_{-0.00076}$ & -- & -- & $0.26^{+0.09}_{-0.12}$, $0.22^{+0.21}_{-0.16}$ & Wong21 \\
$87.2^{+1.6}_{-1.6}$ & $3.630^{+0.051}_{-0.051}$ & $0.0769^{+0.0004}_{-0.0004}$ & -- & -- & $0.319(1)^{\rm {b}}$, $0.227(3)^{\rm {b}}$ & vEssen21 \\
$-$ & $3.43^{+0.15}_{-0.09}$ & $0.0775^{+0.0007}_{-0.0009}$ & -- & -- & $0.42^{+0.10}_{-0.12}$, $-0.05^{+0.19}_{-0.14}$ & Patel22 \\
\multicolumn{7}{c}{KELT-16 b} \\
$82.87^{+0.70}_{-0.62}$ & $3.157^{+0.037}_{-0.035}$ & $0.10961^{+0.00065}_{-0.00065}$ & $0.49^{+0.12}_{-0.13}$, $0.10^{+0.18}_{-0.17}$ & $0.41^{+0.07}_{-0.07}$, $0.07^{+0.13}_{-0.13}$  & $0.40^{+0.10}_{-0.09}$, $0.03^{+0.15}_{-0.15}$  & this paper \\
$84.5^{+2.0}_{-1.4}$ & $3.238^{+0.084}_{-0.075}$ & $0.1076^{+0.0010}_{-0.0010}$ & $0.29^{\rm {a}}$, $0.32^{\rm {a}}$ & $0.36^{\rm {a}}$, $0.29^{\rm {a}}$ & -- & Mac18 \\
$84.8^{+3.0}_{-3.3}$ & $3.21^{+0.10}_{-0.16}$ & $0.1099^{+0.0021}_{-0.0019}$ & $-$ & $-$  & $0.23^{+0.12}_{-0.13}$, $0.34^{+0.28}_{-0.22}$  & Wong21 \\
$89.72^{+0.25}_{-0.25}$ & $3.319^{+0.022}_{-0.022}$ & $0.10814^{+0.00087}_{-0.00087}$ & $-$ & $-$  & $-$  & Mancini22 \\
\multicolumn{7}{c}{WASP-18 b} \\
$83.97^{+0.31}_{-0.31}$ & $3.487^{+0.020}_{-0.020}$ & $0.09777^{+0.00022}_{-0.00022}$ & -- & -- & $0.294^{+0.027}_{-0.026}$, $0.172^{+0.047}_{-0.050}$ & This paper \\
$84.04^{+0.36}_{-0.38}$ & $3.492^{+0.024}_{-0.025}$ & $0.09776^{+0.00028}_{-0.00027}$ & -- & -- & $0.296^{+0.034}_{-0.034}$, $0.159^{+0.061}_{-0.060}$ & Mac20 \\
$84.88^{+0.33}_{-0.33}$ & $3.562^{+0.022}_{-0.023}$ & $0.09716^{+0.00014}_{-0.00013}$ & -- & -- & $0.219^{\rm {c}}$, $0.313^{\rm {c}}$ & Shporer19 \\
\multicolumn{7}{c}{WASP-103 b} \\
$86.5^{+1.9}_{-1.4}$    & $2.977^{+0.031}_{-0.037}$ & $0.11255^{+0.00076}_{-0.00072}$ & $0.23^{+0.12}_{-0.12}$, $0.30^{+0.18}_{-0.18}$ & $0.35^{+0.06}_{-0.07}$, $0.17^{+0.12}_{-0.12}$  & -- & this paper \\
$87.9^{+1.4}_{-1.7}$ & $2.996^{+0.018}_{-0.033}$ & $0.11204^{+0.00070}_{-0.00070}$ & $0.31^{\rm {a}}$, $0.31^{\rm {a}}$ & $0.36^{\rm {a}}$, $0.30^{\rm {a}}$ &  -- & Mac18 \\
$87.0^{+0.2}_{-0.2}$ & $3.01^{+0.01}_{-0.01}$ & $0.1136^{+0.0005}_{-0.0005}$ & $-$ & $-$ &  -- & Kirk21 \\
\hline                                   
\end{tabular}
\tablefoot{$^{\rm {a}}$ interpolated from the theoretical tables of \citet{2011AA...529A..75C} and varied under the Gaussian penalty of the width of 0.1. $^{\rm {b}}$ calculated from PHOENIX library and kept fixed in modeling. $^{\rm {c}}$ fixed at the theoretical values interpolated from the tables of \citet{2017AA...600A..30C}. Data source: Kirk21 -- \citet{2021AJ....162...34K}, Mac18 -- \citet{2018AcA....68..371M}, Mac20 -- \citet{2020AcA....70....1M}, Mancini22 -- \citet{2022MNRAS.509.1447M}, Patel22 -- \citet{2022AJ....163..228P}, Shporer19 -- \citet{2019AJ....157..178S}, vEssen21 -- \citet{2021AA...648A..71V}, the circular-orbit solution, Wong21 -- \citet{2021AJ....162..127W}.}
\end{table*}

\subsection{Transit timing}\label{SSect:Results.Timing}

Transit timing analysis was performed following the procedure described in \citet{2018AcA....68..371M}. In brief words: Linear transit ephemerides were refined using the updated sets of mid-transit times. We used the new mid-transit times, which are collected in Table~\ref{tab.TTimes} in Appendix~\ref{App:1}, together with those compiled in \citet{2018AcA....68..371M} and \citet{2020AcA....70....1M}. For homogeneity, we also redetermined the mid-transit times using publicly available light curves reported in the literature since then. Thus, we enhanced our timing data sets with 27 measurements based on data published by \citet{2022MNRAS.509.1447M} for KELT-16~b and with 12 measurements from \citet{2022AA...657A..52B} for WASP-103~b. The de-trended light curves acquired with the CHaracterising ExOplanet Satellite (CHEOPS) were taken in the latter case. The redetermined transit times are also listed in Table~\ref{tab.TTimes} in Appendix~\ref{App:1}. Finally, we used a compilation of 38 mid-transit times for HAT-P-23~b, 39 for KELT-1~b, 103 for KELT-16~b, 125 for WASP-18~b, and 51 for WASP-103~b.  

\begin{figure*}
	\includegraphics[width=2\columnwidth]{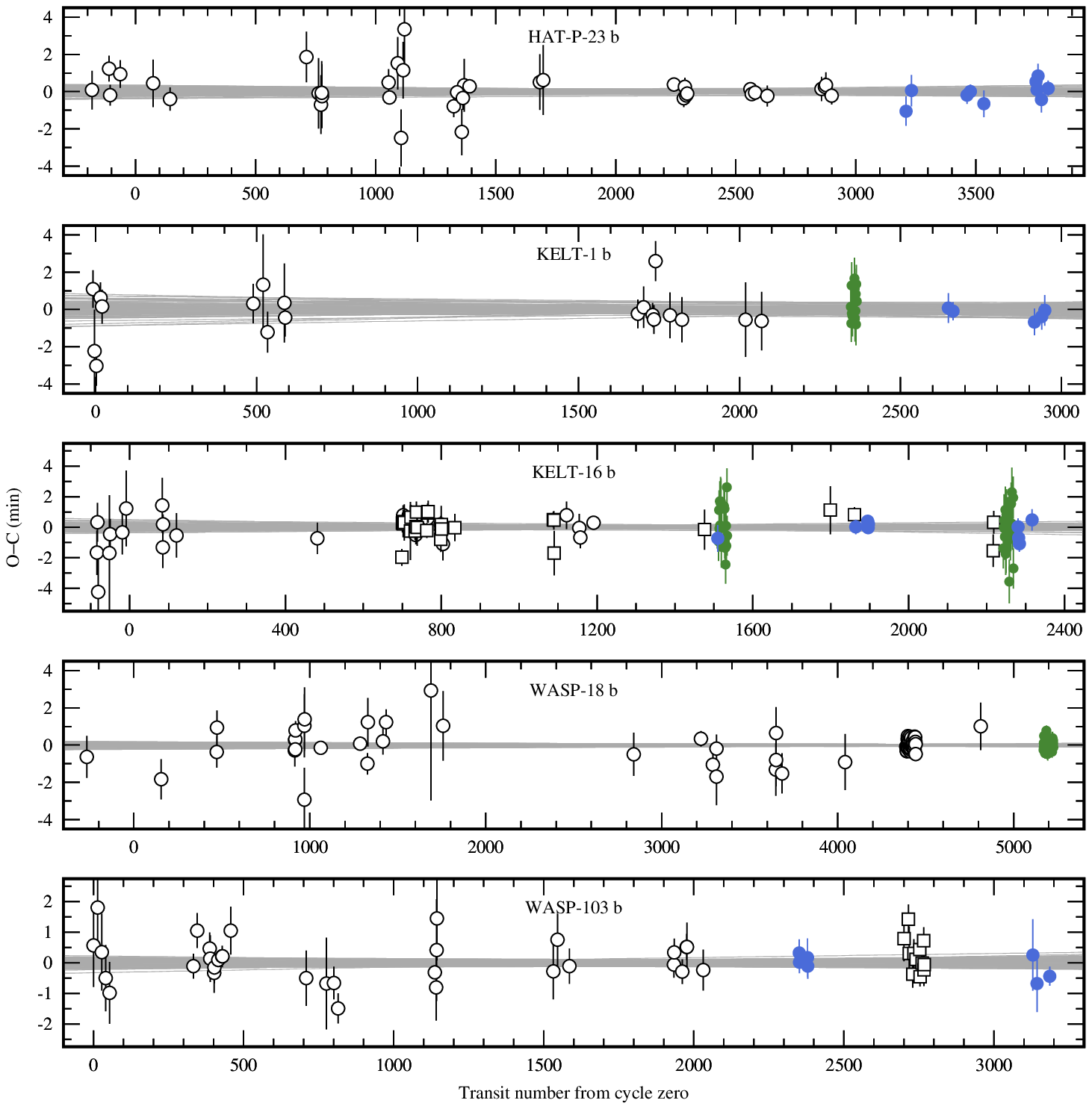}
    \caption{Transit timing residuals for the planets of our sample. The open circles mark the data compiled in \citet{2018AcA....68..371M} and \citet{2020AcA....70....1M}. Open squares are the new literature mid-points redetermined in this study (for Kelt-16~b and WASP-103~b). The filled dots are the new determinations reported in this paper: the green and blue points come from TESS and ground-based photometry, respectively. The uncertainties of the refined ephemerides are illustrated with bunches of 100 lines, each drawn from the Markov chains.}
    \label{fig:ttres}
\end{figure*}

The transit timing datasets were used to refine the linear ephemerides in the form
\begin{equation}
  T_{\rm mid }(E) = T_0 + P_{\rm orb} \cdot E \, , \;
\end{equation}
where $E$ is the transit number counted from the reference epoch $T_0$, taken from the discovery paper. The best-fitting parameters and their $1\sigma$ were derived with the MCMC technique based on 100 chains, each of which was $10^4$ steps long, and the first 1000 trials were discarded. Then, the procedure was repeated for a trial quadratic ephemeris in the form
\begin{equation}
 T_{\rm{mid}}= T_0 + P_{\rm{orb}} \cdot E + \frac{1}{2} \frac{{\rm d} P_{\rm{orb}}}{{\rm d} E} \cdot E^2 \, , \;
\end{equation}
where ${{\rm d} P_{\rm{orb}}}/{{\rm d} E}$ is the change of the orbital period between succeeding transits. The Bayesian information criterion (BIC) values were calculated to assess the preferred model. The results are collected in Table~\ref{tab.Efem}. The values of the quadratic terms were found to be consistent with zero within 0.3--1.8 $\sigma$. The values of BIC also speak in favour of the linear ephemerides for all planets in our sample. The residuals against the refined linear ephemerides are shown in Fig.~\ref{fig:ttres}.

%------------------ TABLE Efem
\begin{table*}[h]
\caption{Parameters of the refined transit ephemerides.} 
\label{tab.Efem}      
\centering                  
\begin{tabular}{l c c c c c c c}      
\hline\hline                
Planet & $T_0$ ($\rm{BJD_{TDB}}$)  & $P_{\rm orb}$ (d) & $\frac{{\rm d} P_{\rm{orb}}}{{\rm d} E}$ $(\cdot 10^{-10})$ & $N_{\rm{dof}}$ & $\chi^2$ & BIC & $Q'_* >$ \\
\hline
\multicolumn{8}{c}{linear ephemerides} \\
HAT-P-23 b & 2454852.26528(13)  & 1.212886436(52) & $-$ &  46 &  32.6 &  40.3 & $-$ \\
KELT-1 b   & 2455909.29304(25)  & 1.21749391(12)  & $-$ &  36 &  29.2 &  36.4 & $-$ \\
KELT-16 b  & 2457247.24798(13)  & 0.968992922(82) & $-$ & 101 & 117.1 & 126.4 & $-$ \\
WASP-18 b  & 2454221.481843(74) & 0.941452411(16) & $-$ & 123 & 104.6 & 114.3 & $-$ \\
WASP-103 b & 2456459.599285(77) & 0.925545423(42) & $-$ &  49 &  54.7 &  62.6 & $-$ \\
\multicolumn{8}{c}{trial quadratic ephemerides} \\
HAT-P-23 b & 2454852.26545(18)  & 1.21288621(17)  & $ 1.14\pm0.82$ &  45 &  30.7 &  40.3 & $2.76^{+0.21}_{-0.21}\cdot 10^{6}$ \\
KELT-1 b   & 2455909.29291(30)  & 1.21749427(44)  & $-2.5\pm2.9$   &  35 &  28.6 &  39.5 & $2.33^{+0.36}_{-0.38}\cdot 10^{6}$ \\
KELT-16 b  & 2457247.24765(24)  & 0.96899357(40)  & $-5.2\pm2.9$   & 100 & 114.3 & 128.2 & $2.95^{+0.23}_{-0.23}\cdot 10^{5}$ \\
WASP-18 b  & 2454221.48181(12)  & 0.94145244(9)   & $-0.09\pm0.28$ & 122 & 104.5 & 119.0 & $1.09^{+0.04}_{-0.04}\cdot 10^{7}$ \\
WASP-103 b & 2456459.59941(13)  & 0.92554517(21)  & $ 1.6\pm1.3$   &  48 &  53.2 &  65.0 & $3.74^{+0.28}_{-0.31}\cdot 10^{6}$ \\
\hline                                   
\end{tabular}
\tablefoot{Uncertainties of $T_0$ and $P_{\rm orb}$ are given in the concise notation. $N_{\rm{dof}}$ is the number of degrees of freedom. $Q'_* >$ is the the lower constraint at the 95\% confidence level.}
\end{table*}
%------------------ END OF TABLE

Since no statistically significant change of $P_{\rm orb}$ was detected, the 5th percentile of the posterior probability distribution of $\frac{d P_{\rm{orb}}}{d E}$ was used to place the lower constraint on $Q'_*$ at the 95\% confidence level. We followed Eq.~(5) from \citet{2018AcA....68..371M}. The values of the planet-to-star mass ratio ${M_{\rm b}}/{M_{\star}}$ were calculated using the radial velocity amplitudes of the orbital motion $K_{\rm{b}}$ following the formula:
\begin{equation}
 \frac{M_{\rm b}}{M_{\star}} = 4.694 \cdot 10^{6} \cdot K_{\rm{b}} P_{\rm{b}}^{1/3} M_{\star}^{-1/3} (\sin{i_{\rm{b}}})^{-1} \, , \;
\end{equation}
where $K_{\rm{b}}$ is in m~s$^{-1}$, $P_{\rm{b}}$ is in days, and $M_{\star}$ is in $M_{\odot}$. The values of $P_{\rm{b}}$ and $i_{\rm{b}}$ were taken from this study. The values of $K_{\rm{b}}$ and $M_{\star}$ come from the most recent redeterminations, which are available in the literature; they are collected together with the references in Table~\ref{tab.LitPars}. The errors of the individual parameters were used to estimate the uncertainty for the constraint on $Q'_*$. The results are given in Table~\ref{tab.Efem}.

%------------------ TABLE LitPars
\begin{table*}[h]
\caption{Literature parameters of the systems under investigation.} 
\label{tab.LitPars}      
\centering                  
\begin{tabular}{l c l c l}      
\hline\hline                
System & $K_{\rm{b}}$ (m~s$^{-1}$)  & source & $M_{\star}$ ($M_{\odot}$) & source \\
\hline
HAT-P-23 & $ 368.5 \pm 17.6$ & \citet{2015AA...577A..54C} & $1.104 \pm 0.047$ & \citet{2015AA...577A..54C} \\
KELT-1   & $4239   \pm 52  $ & \citet{2012ApJ...761..123S} & $1.34 \pm 0.08$   & \citet{2021AA...648A..71V} \\
KELT-16  & $ 494   \pm 25  $ & \citet{2017AJ....153...97O} & $1.195 \pm 0.044$ & \citet{2022MNRAS.509.1447M} \\
WASP-18  & $1813.9 \pm  2.4$ & \citet{2020AcA....70....1M} & $1.294^{+0.063}_{-0.061}$ & \citet{2020AA...636A..98C} \\
WASP-103 & $ 268   \pm 14  $ & \citet{2022AA...657A..52B} & $1.204 \pm 0.046$ & \citet{2022AA...657A..52B} \\
\hline                                   
\end{tabular}
\end{table*}
%------------------ END OF TABLE

\section{Discussion}\label{Sect:Discussion}

For HAT-P-23, \citet{2018AcA....68..371M} placed the lower constraints on $Q'_{\star}$ equal to $5.6 \cdot 10^5$. Then, it was refined to $(6.4 \pm 1.9) \cdot 10^5$ by \citet{2020AJ....159..150P}. Our constraint of $Q'_{\star} > (2.76\pm0.21)\cdot 10^{6}$ is tighter by a factor of $\approx$4. More recently, \citet{2022MNRAS.tmp..595B} have reported $3.8 \cdot 10^6$ but at a higher confidence level of 99\%. For WASP-18~b, precise transit timing observations span over 13 years, making the system the most sensitive probe in the tidal dissipation studies. In the previous study, we eliminated the values of $Q'_{\star}$ lower than $3.9 \cdot 10^{6}$ \citep{2020AcA....70....1M}. In this study, we push this constraint to $Q'_{\star} > (1.09 \pm 0.04) \cdot 10^{7}$. In the case of the WASP-103 system, the results of \citet{2018AcA....68..371M}, \citet{2020AJ....159..150P}, and \citet{2022AA...657A..52B} showed that the orbital period of the giant planet could increase. The presence of a third body in the system or apsidal precession was postulated to explain that finding. Because of that apparently positive derivative of $P_{\rm{orb}}$, \citet{2018AcA....68..371M} and \citet{2022AA...657A..52B} could place constraints on $Q'_{\star}$ only with higher confidence levels: $10^6$ at 99.96\% and $1.6 \cdot 10^6$ at 99.7\%, respectively. \citet{2020AJ....159..150P} reported a rather weak constraint with $Q'_{\star} > (1.1 \pm 0.1) \cdot 10^{5}$ at 95\% confidence. However, the statistical significance of the reported positive period derivative appears to decrease as the span of observations widens. Our new observations allowed us to place the tighter constraint of $Q'_{\star} > 3.74^{+0.28}_{-0.31}\cdot 10^{6}$ with no statistically significant change in the orbital period of WASP-103~b.

If the mechanism of internal gravity wave dissipation operated in the host stars of these three systems, the orbital period shorting could likely be detected in current transit timing data. For HAT-P-23, WASP-18, and WASP-103, the predicted values of $Q'_{\rm {IGW}}$ are $3.5 \cdot 10^5$, $2.6 \cdot 10^6$, and $4 \cdot 10^5$, respectively \citep{2020MNRAS.498.2270B}. Our empirical limits on $Q'_{\star}$ are greater up to one order of magnitude. As predicted by models of \citet{2020MNRAS.498.2270B}, wave breaking does not happen in these stars, preventing their planets from rapid orbital decay.

KELT-16~b has the shortest observational coverage among our sample objects, which results in the weakest constraint on $Q'_{\star}$. \citet{2018AcA....68..371M} and \citet{2020AJ....159..150P} reported the consistent values of $1.1\cdot 10^{5}$ and $(0.9\pm0.2) \cdot 10^{5}$, respectively. More recently, \citet{2022MNRAS.509.1447M} used the TESS data from Sector 15 and the additional ground-based observations to push this constraint to $(2.2\pm0.4) \cdot 10^{5}$. Our timing dataset yields a tighter constraint of $(2.95 \pm 0.23)\cdot 10^{5}$. This results is still too weak to address tidal boosting because the theoretical prediction is $Q'_{\rm {IGW}} = 7 \cdot 10^5$ \citep{2020MNRAS.498.2270B}. As the mass of the star is $\approx 1.2$ $M_{\odot}$ \citep{2022MNRAS.509.1447M}, this tidal boost is, however, improbable.

The KELT-1 system was recognised as the best candidate for studying the star-planet tidal interaction \citep{2018AcA....68..371M} unless the stellar spin is synchronised to the orbital period. The rotation period of the host star was found to be equal to $P_{\star} = 1.33 \pm 0.06$ d from the measured projected rotation velocity \citep{2012ApJ...761..123S} and $P_{\star} = 1.52 \pm 0.29$ d from photometric variations \citep{2021AA...648A..71V}. Both results differ by more than 1$\sigma$ from the orbital period of KELT-1~b, which is $1.22$ d. Thus, the spin and orbit synchronisation may still be ongoing. The tidal period in the KELT-1 system is $\approx 7$ days and is longer than half the host star's rotation. In such a configuration, tidal dissipation might be enhanced by interactions of inertial waves and turbulent motions in a convective layer of the star. Since an analytic formula that estimates the theoretical value of $Q'_{\rm {IW}}$ is not available we utilised Figs. 4 and 6 of \citet{2020MNRAS.498.2270B} to find $Q'_{\rm {IW}} \approx 2.6 \cdot 10^6$ for KELT-1. Our empirical constraint of $2.33^{+0.36}_{-0.38}\cdot 10^{6}$ does not reject this value. We also calculated $Q'_{\rm {IGW}}$ using Equation (44) from \citet{2020MNRAS.498.2270B}. Its value of $\approx 5 \cdot 10^8$ is well beyond the detection limit of current and near-future transit timing observations.

\section{Conclusions}\label{Sect:Conclusions}

Our transit timing data show that tidal dissipation is not boosted by breaking internal gravity waves in HAT-P-23, WASP-18, and WASP-103 host stars. This negative result is in line with the models of stellar interiors of stars with masses above $1.1$ $M_{\odot}$ for which convective cores prevent the waves from reaching the stellar centres and breaking. For KELT-16, the span of observations is not long enough to verify the theoretical predictions yet. Precise transit timing in the following years will allow us to probe the regime of dynamical tides in that system.

The KELT-1 system was found not to be a favourable laboratory for studying tidal dissipation if the equilibrium tides or internal gravity wave mechanisms are considered. However, it might become a unique tool for probing the dissipative tidal interactions of inertial waves in convective layers. Further observations will help us to explore that scenario.

\begin{acknowledgements}
We thank the anonymous referee for the comments that improved the quality of this paper. GM acknowledges the financial support from the National Science Centre, Poland through grant no. 2016/23/B/ST9/00579. MF and PJA acknowledge financial support from grant PID2019-109522GB-C5X/AEI/10.13039/501100011033 of the Spanish Ministry of Science and Innovation (MICINN) and from grant P20-00737 of the Andalusian Government program PAIDI 2020. MF, AS, and PJA acknowledge financial support from the State Agency for Research of the Spanish MCIU through the \textit{Center of Excellence Severo Ochoa} award to the Instituto de Astrof\'{\i}sica de Andaluc\'{\i}a (SEV-2017-0709). MM and RB acknowledge the support of the DFG priority programme SPP 1992 \textit{Exploring the Diversity of Extrasolar Planets} (NE 515/58-1 and MU 2695/27-1). This paper includes data collected with the TESS mission, obtained from the MAST data archive at the Space Telescope Science Institute (STScI). Funding for the TESS mission is provided by the NASA Explorer Program. STScI is operated by the Association of Universities for Research in Astronomy, Inc., under NASA contract NAS 5-26555. This research made use of Lightkurve, a Python package for Kepler and TESS data analysis \citep{2018ascl.soft12013L}. This research has made use of the SIMBAD database and the VizieR catalogue access tool, operated at CDS, Strasbourg, France, and NASA's Astrophysics Data System Bibliographic Services.
\end{acknowledgements}

\bibliographystyle{aa} % style aa.bst 
\bibliography{lit} % your references Yourfile.bib

\begin{appendix}
\section{New mid-transit times}\label{App:1}

The transit mid-points obtained from the new transit light curves and those that are available in the literature are collected in Table~\ref{tab.TTimes}. They were combined with timing datasets from \citet{2018AcA....68..371M} and \citet{2020AcA....70....1M} for transit timing studies based on homogeneously determined data.

%------------------ TABLE TTimes
\begin{table*}[h]
\caption{Mid-transit times reported in this paper.} 
\label{tab.TTimes}      
\centering                  
\begin{tabular}{c c c l}      
\hline\hline                
$T_{\rm{mid}}$ $(\rm{BJD_{TDB}})$ & $+\sigma$ (d) & $-\sigma$ (d)  &  Light curve\\
\hline
\multicolumn{4}{c}{HAT-P-23 b} \\
2458744.417125 & 0.000547 & 0.000544 & Jena \\ 
2458772.314281 & 0.000581 & 0.000583 & Jena \\ 
2459052.490879 & 0.000329 & 0.000347 & Trebur \\ 
2459069.471419 & 0.000261 & 0.000264 & OSN150 \\ 
2459137.392604 & 0.000495 & 0.000491 & OSN150 \\ 
2459400.589783 & 0.000293 & 0.000295 & OSN90 \\ 
2459405.441030 & 0.000300 & 0.000308 & Rozhen \\ 
2459411.505979 & 0.000443 & 0.000447 & OSN90 \\ 
2459428.485508 & 0.000483 & 0.000457 & OSN90 \\ 
2459462.446747 & 0.000290 & 0.000285 & OSN150 \\
\multicolumn{4}{c}{KELT-1 b} \\
... & ... & ... & ... \\ 
\hline                                   
\end{tabular}
\end{table*}
%------------------ END OF TABLE

\end{appendix}
\end{document}